# A note on structured means analysis for a single group

André Beauducel[1]

October 3$^{rd}$, 2015

**Abstract**

The calculation of common factor means in structured means analysis (SMM) is considered. The SMM equations imply that the unique factors are defined as having zero means. It was shown within the one factor solution that this definition implies larger absolute common factor loadings to co-occur with larger absolute expectations of the observed variables in the single group case. This result was illustrated by means of a small simulation study. It is argued that the proportionality of factor loadings and observed means should be critically examined in the context of SMM. It is recommended that researchers should freely estimate the observed expectations, whenever, the proportionality of loadings and observed means does not make sense. It was also shown that for a given size of observed expectations, smaller common factor loadings result in larger common factor mean estimates. It should therefore be checked whether variables with very small factor loadings occur when SMM results in very large absolute common factor means.

---

[1] Dr. André Beauducel, Institute of Psychology, University of Bonn, Kaiser-Karl-Ring 9, 53111 Bonn, Germany, email: beauducel@uni-bonn.de

Sörbom (1974) proposed estimates of factor means for confirmatory factor analysis (CFA) and structural equation modeling (SEM). The method was called structured means modeling (SMM) and is widely used in order to test hypotheses on group differences (Cole, Maxwell, Arvey & Salas, 1993; Hancock, Lawrence, & Nevitt, 2000). Several methodological developments have been presented in order to improve SMM, especially for the testing of group differences (Hancock et al., 2000; Whittaker, 2013). However, the estimation of common factor means might also be considered in the single group case. The present note therefore refers to SMM for a single group of participants.

### Definitions

Sörbom (1974) presented the original equations for SMM for multiple groups. Accordingly, the factor model for latent mean analysis for the $g$th group can be written as

$$\mathbf{x}_g = \boldsymbol{\mu} + \boldsymbol{\Lambda}_g \boldsymbol{\xi}_g + \boldsymbol{\varepsilon}_g, \quad with \ g = 1, 2, ..., m, \tag{1}$$

where $\mathbf{x}_g$ is a vector of $p$ observed variables, $\boldsymbol{\mu}$ is the constant (intercept) of the observed variables, $\boldsymbol{\Lambda}_g$ is a $p$ x $q$ matrix of factor loadings, $\boldsymbol{\xi}_g$ is a random vector of common factors, and $\boldsymbol{\varepsilon}_g$ is a random vector of unique factors. Equation 1 follows Sörbom's (1974) notation. The usual assumptions of the factor model are that the covariances of the unique factors are a diagonal matrix of unique factor variances (Cov[$\boldsymbol{\varepsilon}_g$, $\boldsymbol{\varepsilon}_g$] = $\boldsymbol{\Psi}_g^2$) and that the covariances of the common and unique factors are zero (Cov[$\boldsymbol{\xi}_g$, $\boldsymbol{\varepsilon}_g$] = $\mathbf{0}$).

The expected values of the observed variables for the $g$th group are given by

$$E(\mathbf{x}_g) = \boldsymbol{\mu} + \boldsymbol{\Lambda}_g \boldsymbol{\theta}_g, \tag{2}$$

where $\boldsymbol{\theta}_g$ is the mean vector of the common factors. This is the equation as it is usually presented (Sörbom, 1974; Cole et al., 1993). The expectancy refers to the population of individuals in the $g$th group. It should be noted that Equation 2 implies $E(\boldsymbol{\varepsilon}_g) = \mathbf{0}$, i.e., that the unique factors have a zero expectation. Therefore, the definition of common factor means according to Equation (2) corresponds to the component model. This follows from the fact that the expectations of the observed variables are calculated as a linear combination from the common factor loadings and that the loadings of the unique factors are not taken into account. Accordingly, the common factor mean $\boldsymbol{\theta}_g$ can be calculated from the observed expectations when Equation 2 is transformed into

$$(\boldsymbol{\Lambda}'_g \boldsymbol{\Lambda}_g)^{-1} \boldsymbol{\Lambda}'_g (E(\mathbf{x}_g) - \boldsymbol{\mu}) = \boldsymbol{\theta}_g, \tag{3}$$

where the expectations of the observed variables result from the group specific expectations $E(\mathbf{x}_g)$ and the constant $\boldsymbol{\mu}$.

## The single group case

When there is only a single group, it is impossible to distinguish between the overall constant $\boldsymbol{\mu}$ and the group-specific expectations $E(\mathbf{x}_g)$ of the observed variables. Therefore, only the expectations of the observed variables can be considered, which implies

$$(\boldsymbol{\Lambda}'\boldsymbol{\Lambda})^{-1}\boldsymbol{\Lambda}' E(\mathbf{x}) = \boldsymbol{\theta}, \tag{4}$$

since the group specific index is not necessary for the single group case.

Equation 2 for a single group can be written as

$$E(\mathbf{x}) = \boldsymbol{\Lambda}\boldsymbol{\theta}. \tag{5}$$

Considering a one-factor model yields

$$E(\mathbf{x}) \oslash \boldsymbol{\Lambda} = \mathbf{1}\boldsymbol{\theta}, \tag{6}$$

where "$\oslash$" denotes the Hadamard (elementwise) division and "$\mathbf{1}$" is a $p \times 1$ unit-vector. Equation 6 implies that the ratio of the expectations of observed variables and the corresponding loadings yields $\boldsymbol{\theta}$. It follows that performing conventional SMM for single group data in order to estimate a single common factor model with a common factor mean implies a proportionality of common factor loadings and observed expectations. This effect is also present with more than one common factor, as long as the variables have only a single common factor loading. It might be less pronounced when each variable has more than one factor loading. However, several applications of the SMM are possible where this proportionality might not be reasonable: For example, why should a more easy task with a larger mean number of correct solutions have higher factor loadings than a more difficult task with a smaller mean number of correct solutions?

The model implied by Equations 5 and 6 may fit to the data when floor effects occur for some of the observed variables: Floor effects typically result in smaller variances of the observed variables with smaller means which typically result in smaller inter-correlations of these variables. In such a context of variables, those variables with larger means will have larger factor loadings which will correspond to Equations 5 and 6. In contrast, a ceiling effect might cause the variables with higher means to have smaller variances so that they might have smaller factor loadings. However, a calculation of common factor means that is more compatible with a floor effect than with a ceiling effect may introduce an arbitrary bias into single group SSM.

As an illustration for the implications of Equations 5 and 6, a population one-factor model (Model 1) with five observed variables and common factor loadings of .30, .40, .50, .60, and .70 with expectations of the observed variables of 3, 4, 5, 6, and 7 was defined. Model 1 can be directly derived from Equation 5 when a common factor mean of 10 is assumed. Model 2 had exactly the same common factor loadings and a reversed order of expectations of the observed variables (7, 6, 5, 4, and 3; see Table 1). Thus, Model 2 cannot be derived from Equation 5 since there is no constant that can be entered as a common factor mean in order to get the respective expectations. For these two population models a small simulation study based on samples with 150, 300, and 900 cases based on 2,000 replications for each model and sample size was conducted (see Table 1). One-factor models were specified, common factor loadings were freely estimated and the factor variances were fixed at one. In order to be compatible with Equation 5, the intercepts of the observed variables were fixed to zero.

As expected, Model 1 fits perfectly to the data with a mean $\chi^2$-value that is close to the degrees of freedom (see Table 1) and the estimated common factor mean is close to the value used for model generation. In contrast, Model 2 does not fit to the data since the mean $\chi^2$-value is much larger than the degrees of freedom. Although the degree of misfit of Model 2 is strongly affected by sample size, the averaged model parameters are very similar for all samples sizes (see Table 1). Moreover, the relative size of the estimated loadings is reversed for Model 2, so that the loadings are proportional to the observed variables expectations (although this was not the case in the population model). Overall, these results illustrate the consequences of common factor mean estimation in accordance with Equation 5. Model 1, which is based on a proportionality of common factor loadings and observed expectations has a considerably superior fit than Model 2 which is based on anti-proportional factor loadings and observed expectations.

**Table 1.**
**Factor loadings and factor means for two population models in the single group case**

|  | Population data | | | |
|---|---|---|---|---|
| Item | F1 (Model 1) | | F1 (Model 2) | |
|  | Λ | μ | Λ | μ |
| x1 | .30 | 3.00 | .30 | 7.00 |
| x2 | .40 | 4.00 | .40 | 6.00 |
| x3 | .50 | 5.00 | .50 | 5.00 |
| x4 | .60 | 6.00 | .60 | 4.00 |
| x5 | .70 | 7.00 | .70 | 3.00 |
|  | 2,000 samples, N = 900 | | | |
| Item | F1 (Model 1) | | F1 (Model 2) | |
|  | $\hat{\Lambda}$ | μ* | $\hat{\Lambda}$ | μ* |
| x1 | .30 (0.01) | 0.00 | .56 (0.03) | 0.00 |
| x2 | .40 (0.02) | 0.00 | .48 (0.03) | 0.00 |
| x3 | .50 (0.02) | 0.00 | .40 (0.02) | 0.00 |
| x4 | .60 (0.03) | 0.00 | .32 (0.02) | 0.00 |
| x5 | .70 (0.03) | 0.00 | .24 (0.01) | 0.00 |
| Factor mean | 10.04 (0.43) | | 12.49 (0.66) | |
| Model fit | $\chi^2 = 9.15$ (4.26); $df = 9$ | | $\chi^2 = 126.82$ (22.88); $df = 9$ | |
|  | 2,000 samples, N = 300 | | | |
| Item | F1 (Model 1) | | F1 (Model 2) | |
| x1 | .30 (0.02) | 0.00 | .56 (0.05) | 0.00 |
| x2 | .40 (0.03) | 0.00 | .48 (0.04) | 0.00 |
| x3 | .50 (0.04) | 0.00 | .40 (0.04) | 0.00 |
| x4 | .60 (0.05) | 0.00 | .32 (0.03) | 0.00 |
| x5 | .70 (0.05) | 0.00 | .24 (0.02) | 0.00 |
| Factor mean | 10.10 (0.79) | | 12.59 (1.21) | |
| Model fit | $\chi^2 = 9.01$ (4.28); $df = 9$ | | $\chi^2 = 48.47$ (13.42); $df = 9$ | |
|  | 2,000 samples, N = 150 | | | |
| Item | F1 (Model 1) | | F1 (Model 2) | |
| x1 | .30 (0.03) | 0.00 | .56 (0.07) | 0.00 |
| x2 | .40 (0.04) | 0.00 | .48 (0.06) | 0.00 |
| x3 | .49 (0.05) | 0.00 | .40 (0.05) | 0.00 |
| x4 | .59 (0.06) | 0.00 | .32 (0.04) | 0.00 |
| x5 | .69 (0.07) | 0.00 | .24 (0.03) | 0.00 |
| Factor mean | 10.25 (1.16) | | 12.83 (1.89) | |
| Model fit | $\chi^2 = 9.16$ (4.32); $df = 9$ | | $\chi^2 = 28.46$ (9.99); $df = 9$ | |

*Note.* * Parameters were fixed for model specification. Standard deviations are given in brackets.

Note that the simple way out of the problem is not to freely estimate the common factor mean and all the observed variables means, because such a model would not be identified. It is, however, possible to fix one observed variable mean per factor and to estimate the common factor mean on this basis. This strategy would be incompatible with Equation 5 and it introduces the choice of the observed variables mean to be fixed. When a single variables mean is fixed to zero, the common factor mean is calculated on the basis of the empirical mean and the loading of this variable. When the expectations of all observed variables are equal, it follows from Equation 5 that choosing a variable with a small common factor loading will lead to a larger common factor mean than choosing a variable with a large common factor loading. When there is, moreover, variation of the loading sizes on one factor, the choice of the variable with the fixed mean will affect the common factor mean. For example, in Model 2 of the simulation study (Table 1), fixing the mean of the variable with the smallest loading (x1) to zero results in an averaged common factor mean of 23.82 (SD = 3.90) in the simulation based on 900 cases. In contrast, in the same condition, fixing the mean of variable with the largest loading (x5) to zero results in a common factor mean of 4.32 (SD = 0.37). Thus, the modeling strategy of fixing a single mean introduces considerably arbitrariness into the resulting common factor mean.

The relation between common factor loadings and the resulting common factor mean can be more generally described. Theorem 1 shows the consequences of estimating factor means according to Equation 6, when factor models based on observed variables with the same $E(\mathbf{x})$ but different $\mathbf{\Lambda}$ are compared. Consider a one-factor model with equal loadings on the common factor, so that $\mathbf{\Lambda} = w\mathbf{1}$, where $\mathbf{1}$ is a $p \times 1$ unit-vector, and $w$ is a scalar.

**Theorem 1.** $w \to 0$ *implies* $abs(\mathbf{\theta}) \to \infty$ *for* $E(\mathbf{x}) \neq \mathbf{0}$,

with *"abs"* denoting absolute values.

*Proof.* Entering $\mathbf{\Lambda} = w\mathbf{1}$ into Equation 6 yields

$$(w\mathbf{1}'\mathbf{1}w)^{-1} w\mathbf{1}' \ E(\mathbf{x}) = p^{-1} w^{-1} \mathbf{1}' \ E(\mathbf{x}) = \mathbf{\theta}. \tag{7}$$

This completes the proof. □

Theorem 1 describes the following relationship for constant $E(\mathbf{x})$: The smaller the absolute size of the common factor loadings, the larger is the absolute size of the common factor mean. On this basis, one might question whether a modeling of factor means according to Equation 5 and 6 is reasonable in the single-group case, since lower factor loadings normally imply that a more careful interpretation of the factor would be reasonable.

## Discussion

The effect of SMM on the descriptive common factor means was investigated for the single-group case. It was found that conventional SMM, which is based on the equations of Sörbom (1974), implies that the expectations of the observed variables are proportional to the absolute size of the common factor loadings. There is, however, no theoretical reason for the expectations of observed variables being proportional to the common factor loadings. In several tests or inventories the relative size of the expectations depends on a more or less arbitrary scale of the variables. In other inventories, the expectations also depend on ceiling effects and floor effects. Interestingly, floor effects implying smaller variances, thereby smaller inter-correlations and factor loadings for variables with smaller expectations are compatible with the assumptions of the SMM, whereas ceiling effects are not. It follows that researchers should check, whether such effects are present in their data, in order to be aware in their effects on SMM.

Moreover, even equal means of the observed variables could be a cause of model misfit, when the factor loadings are unequal. In order to avoid model evaluations that are based on the misfit that is caused by the proportionality assumption of the SMM, researchers should freely estimate some of the intercepts of the observed variables, whenever the proportionality assumption of the SMM does not make sense in the specific context of research. However, this leads to the problem of deciding which intercept has to be freely estimated. Since it was shown that larger absolute common factor means occur for smaller absolute common factor loadings when the observed expectations (intercepts) remain unchanged (Theorem 1), it is recommended to freely estimate the intercepts of the variables with the smallest loadings. This would ensure that the most important observed variables are used in order to estimate the common factor mean and that the common factor mean is not inflated due to the effect of small loadings.